# COOPERATIVE ORIGIN OF LOW-DENSITY DOMAINS IN LIQUID WATER

by

Jeffrey R. Errington¶ §, Pablo G. Debenedetti¶ [*], and Salvatore Torquato‡

¶ Department of Chemical Engineering, Princeton University, Princeton, NJ 08544
§ Current address: Department of Chemical Engineering, State University of New York, Buffalo, NY 14260
‡ Department of Chemistry, Princeton University, Princeton, NJ 08544

## ABSTRACT

We study the size of clusters formed by water molecules possessing large enough tetrahedrality with respect to their nearest neighbors. Using Monte Carlo simulation of the SPC/E model of water, together with a geometric analysis based on Voronoi tessellation, we find that regions of lower density than the bulk are formed by accretion of molecules into clusters exceeding a minimum size. Clusters are predominantly linear objects and become less compact as they grow until they reach a size beyond which further accretion is not accompanied by a density decrease. The results suggest that the formation of "ice-like" regions in liquid water is cooperative.

PACS number: 61.20.-p

[*] Corresponding author. E-mail pdebene@princeton.edu

Water is an unusual liquid [1-4]. At low enough temperatures and pressures, it expands when cooled, becomes less viscous when compressed and more compressible when cooled, and its already large isobaric heat capacity increases sharply upon cooling. Because of its essential role in biology, atmospheric phenomena, geology, and technology, there is widespread interest in understanding the molecular origin of water's anomalous properties [1].

Perhaps the best-known anomaly of liquid water is its ability to expand when cooled at constant pressure (negative thermal expansion). This phenomenon can be understood qualitatively in terms of the formation of localized molecular arrangements having a lower local density than the bulk. Molecules in these transient low-density configurations must, on average, adopt more structured arrangements than the bulk liquid. Such local, transient, structured arrangements are often described as "ice-like". While the soundness of this picture is not in dispute (freezing in water is accompanied by a 7.7 % decrease in density), its details are not well understood. The size, shape, and lifetime of "ice-like" clusters have not, to our knowledge, been determined. Here we investigate a basic geometric aspect of this problem. We ask the question: what is the shape and size of clusters whose density is lower than the bulk density? We use Monte Carlo simulation of the shifted force SPC/E model of water [5, 6] to answer the question.

We first introduce an orientational order parameter $q$ [7, 8] that measures the extent to which a molecule and its four nearest neighbors adopt a tetrahedral arrangement, such as exists in hexagonal ice (Ih):

$$q = 1 - \frac{3}{8} \sum_{j=1}^{3} \sum_{k=j+1}^{4} \left( \cos \psi_{jk} + \frac{1}{3} \right)^2 \tag{1}$$

In the above equation, $\psi_{jk}$ is the angle formed by the lines joining the oxygen atom of a given molecule and those of its nearest neighbors $j$ and $k$ ($= 4$). If a molecule is located at the center of a regular tetrahedron whose vertices are occupied by its four nearest neighbors, $\cos \psi_{jk} = -1/3$. Thus, in a perfect tetrahedral network, $q = 1$. If, on the other hand, the mutual arrangement of molecules is random, as in an ideal gas, the six angles associated with the central molecule are independent, and the mean value of $q$ vanishes:

$$\langle q \rangle = 1 - \frac{9}{8} \int_0^{\pi} \left( \cos \psi + \frac{1}{3} \right)^2 \sin \psi \, d\psi = 0 \tag{2}$$

We study clusters formed by molecules whose orientational order parameter $q$ exceeds a certain value $q_c$. Specifically, if two molecules have $q$-values larger than $q_c$ and are separated by a distance smaller than $r_c$, they belong to the same cluster. At any given thermodynamic state point, we chose the location of the first minimum in the O-O pair correlation function as the value for $r_c$.

State points studied in this work span the temperature range $220 = T = 300$K, at atmospheric pressure. An initial set of isothermal-isobaric simulations was performed to calculate the densities corresponding to 220, 240, 260, 280 and 300K at atmospheric pressure. The results reported here correspond to subsequent $(N,V,T)$ simulations at these calculated densities (Table 1). This indirect way of imposing isobaric conditions was adopted because the geometric analysis on which this work is based involves the calculation of distances and volumes whose natural fluctuation is greatly amplified by volume fluctuations in $(N,P,T)$ simulations.

Figure 1 shows the cluster size distribution at $T = 240$K for various choices of $q_c$. The curves for $q_c = 0.7$ and 0.75 indicate the presence of a large, system-spanning cluster (the number of molecules in the simulation cell is 256). As $q_c$ increases, however, smaller clusters are seen. They are formed by molecules that adopt a progressively more tetrahedral arrangement with their four nearest neighbors. It is on these clusters composed of individual molecules each having a high degree of tetrahedrality that we focus our attention.

Figure 2 shows the average volume per molecule in a cluster as a function of the cluster size (number of molecules in the cluster), at 220, 240, and 260K (the temperature of maximum density is 240K; see Table 1). The cluster volume is the sum of the Voronoi volumes of the molecules belonging to the cluster [9, 10]. Normalization by the number of molecules yields the average volume per molecule in the cluster. The horizontal line corresponds to the bulk specific volume (note that this is minimum at 240K). In each case, only the high-$q_c$ curves (0.85, 0.9) rise above the bulk specific volume. Significantly, we find that only those clusters that exceed a threshold size attain a specific volume larger than the bulk. Furthermore, we see that the specific volume of small clusters and individual molecules that adopt highly tetrahedral arrangements with their four nearest neighbors is *smaller* than the bulk value. This means that the formation of "ice-like" regions having a density lower than the bulk is cooperative, and involves many molecules.

At 220K, the specific volume for clusters composed of molecules with $q_c = 0.85$ reaches a plateau value. This means that clusters become looser when they grow until they reach a critical size, beyond which further accretion is not accompanied by a density decrease. This saturation effect can be seen for all values of $q_c$ examined, whenever it is possible to observe large enough clusters. In the $q_c = 0.9$ case, the curvature at 220K is consistent with volume saturation, but we did not observe large enough clusters of such highly structured nature for this effect to occur. At higher temperatures, we could only observe the initial growth of the $q_c = 0.9$ curve.

Figure 3 shows snapshots of a 15- and a 20-membered water cluster ($q_c = 0.85$, $T = 240$K K, $r = 1$ g/cm$^3$). These are typical of many observations, and show that low-density regions in liquid SPC/E water consist of linear or ramified clusters. We see no evidence of three-dimensional, clathrate-like objects. Pentagonal and hexagonal rings, one of which is shown in Figure 3a, form in sufficiently large clusters. Although hydrogen bonds are not invoked in the definition of a cluster, it can be seen that clusters are in fact spontaneously held together by hydrogen bonds.

Extensive statistics on clusters such as those shown in Figure 3 were collected, and the analysis yielded the results shown in Figures 4, 5, and 6. The quantity plotted in Figure 4 is the number of bonds per cluster normalized by the minimum number of bonds, as a function of the cluster size. Calculations are shown for $T = 220$, 240, 260, 280, and 300K. A bond in this context connects two molecules belonging to the same cluster and separated by a distance less than $r_c$. The minimum number of bonds in a cluster composed of $n$ molecules is ($n$-1), corresponding to a linear arrangement. In accord with the images of Figure 3, it can be seen that the clusters are predominantly linear objects: even 80-membered aggregates have barely 8% more bonds than a strictly linear object of the same number of molecules. Ring statistics at $T = 220$, 240, 260, 280 and 300K are shown in Figures 5 and 6. It can be seen that on average one in every two 30-membered cluster contains a five or six-membered ring, and every 50-membered cluster contains on average one such ring. Smaller, highly strained rings are virtually never seen. Although the number of clusters of a given size depends on temperature, their geometric characteristics show little sensitivity to temperature changes. Six-membered rings have been shown recently to play a key role in the nucleation of ice from supercooled water [11].

There is a substantial literature on so-called mixture models of water [12]. In this approach, molecules are also assigned to one of two categories, often referred to as liquid-like and ice-like. The former have a smaller specific volume and a larger enthalpy. The present analysis implies that these concepts do not apply locally. Molecules with a highly ordered local environment, as measured by their tetrahedrality with respect to their four nearest neighbors, only acquire a larger-than-average specific volume when they accrete onto a sufficiently large cluster of tetrahedrally-arranged molecules. Ice-like character, in other words, is cooperative. It is interesting to note that clusters similar to the ones reported here have been recently found to be responsible for the transitions between potential energy minima in supercooled SPC/E water [13].

X-ray diffraction measurements [14] and computer simulation [15-18] have provided molecular-level insight into water's density maximum. Comparison of the center-of-mass pair correlation function for thermodynamic states with the same density but different temperature revealed an increase in the number of so-called interstitial molecules at the expense of hydrogen-bonded molecules upon heating across the temperature of maximum density (interstitial molecules form closely-packed arrangements with a central molecule and its hydrogen-bond-acceptor neighbors). The present study suggests that molecular-level understanding of the mechanism by which water expands upon cooling remains incomplete. In particular, the formation of low-density domains is cooperative, and cannot be explained solely in terms of structural changes occurring at the nearest- or next-nearest neighbor level.

While the present analysis clarifies the static aspects associated with the formation of low-density regions in cold liquid water, the lifetime of clusters such as those shown in Figure 3 is important and remains to be investigated. It would be also interesting to identify unambiguously the clusters responsible for water's negative thermal expansion. Both these issues are the subject of our ongoing investigations.

**ACKNOWLEDGEMENT**

PGD gratefully acknowledges the support of the U.S. Department of Energy, Division of Chemical Sciences, Geosciences and Biosciences, Office of Basic Energy Science (grant DE-FG02-87ER13714).

**Table 1.** State points examined in this work.

| Temperature (K) | Density[a] (g/cm$^3$) | $r_c$ (Å)[b] |
|---|---|---|
| 220 | 0.9748 (46) | 3.225 |
| 240 | 0.9843 (26) | 3.250 |
| 260 | 0.9797 (34) | 3.270 |
| 280 | 0.9744 (17) | 3.320 |
| 300 | 0.9653 (13) | 3.350 |

[a] Calculated from (*N,P,T*) Mote Carlo simulations using the shifted-force SPC/E model for water

[b] radial distance to the first minimum in the oxygen-oxygen pair correlation function

# Figure Captions

**Figure 1.** The cluster size distribution at $T = 240$ K for various choices of $q_c$: 0.7 (empty circles), 0.75 (empty squares), 0.8 (empty rhombi), 0.85 (empty triangles), 0.9 (filled circles).

**Figure 2.** The average volume per molecule as a function of the cluster size at $T = 220$ (a), 240 (b) and 260K (c), for various choices of $q_c$. The horizontal line is the mean (bulk) volume per molecule for the entire simulation cell.

**Figure 3.** Snapshots of two representative water clusters from a simulation at $T = 240$ K using $q_c = 0.85$. The top and bottom pictures are of a 15- and 20-membered cluster respectively.

**Figure 4.** The number of bonds per cluster normalized by $(n - 1)$, the number of bonds in a linear cluster of that size ($n$ is the number of molecules in the cluster), as a function of the number of molecules in a cluster. The clusters were identified using $q_c = 0.85$ and data were collected at the different temperatures shown.

**Figure 5.** The average number of 5-membered rings per cluster as a function of cluster size. The clusters were identified using $q_c = 0.85$ and data were collected at the different temperatures shown.

**Figure 6.** Identical to Figure 5, but for 6-membered rings.

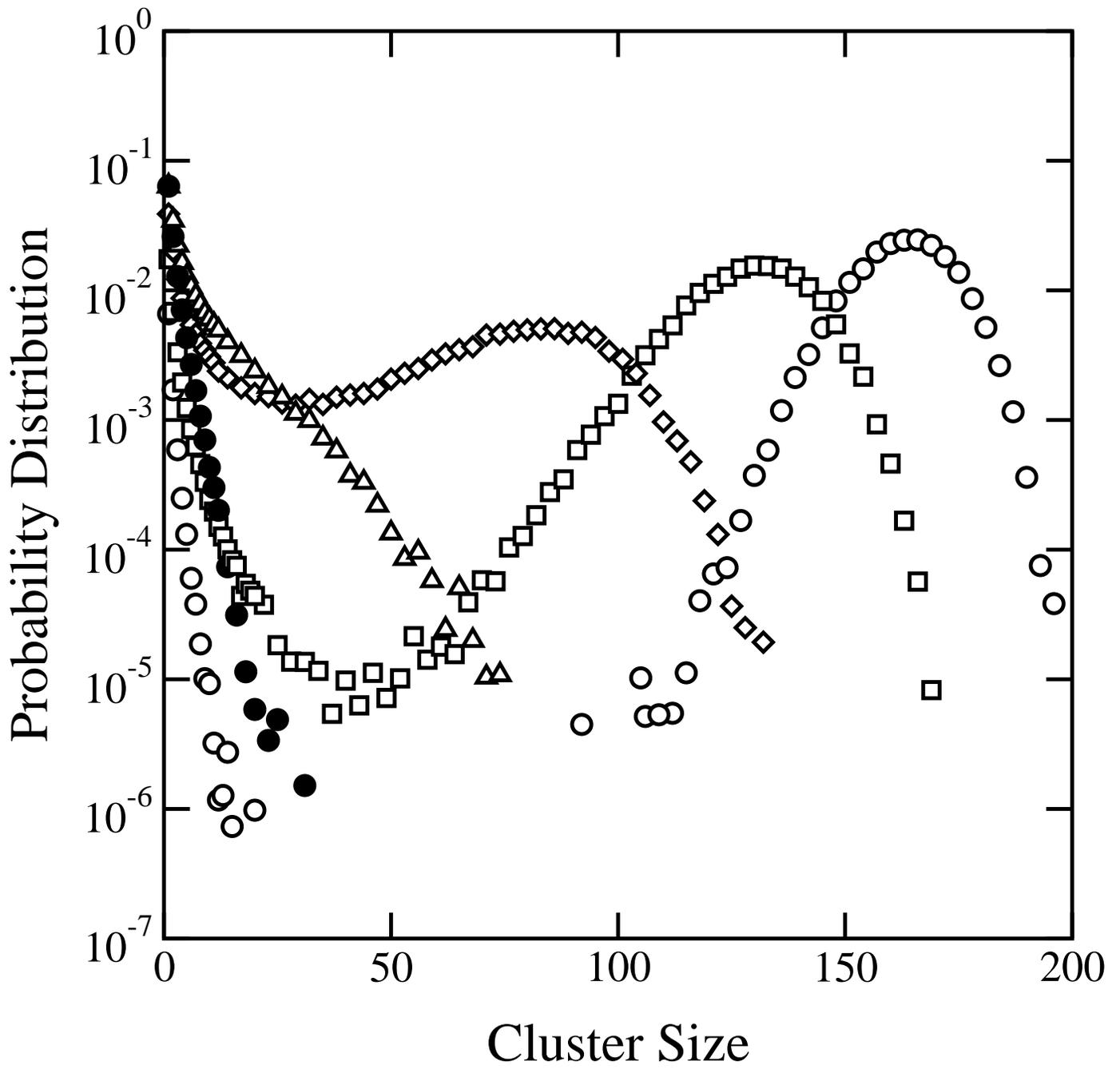

Figure 1. Errington, Debenedetti, and Torquato

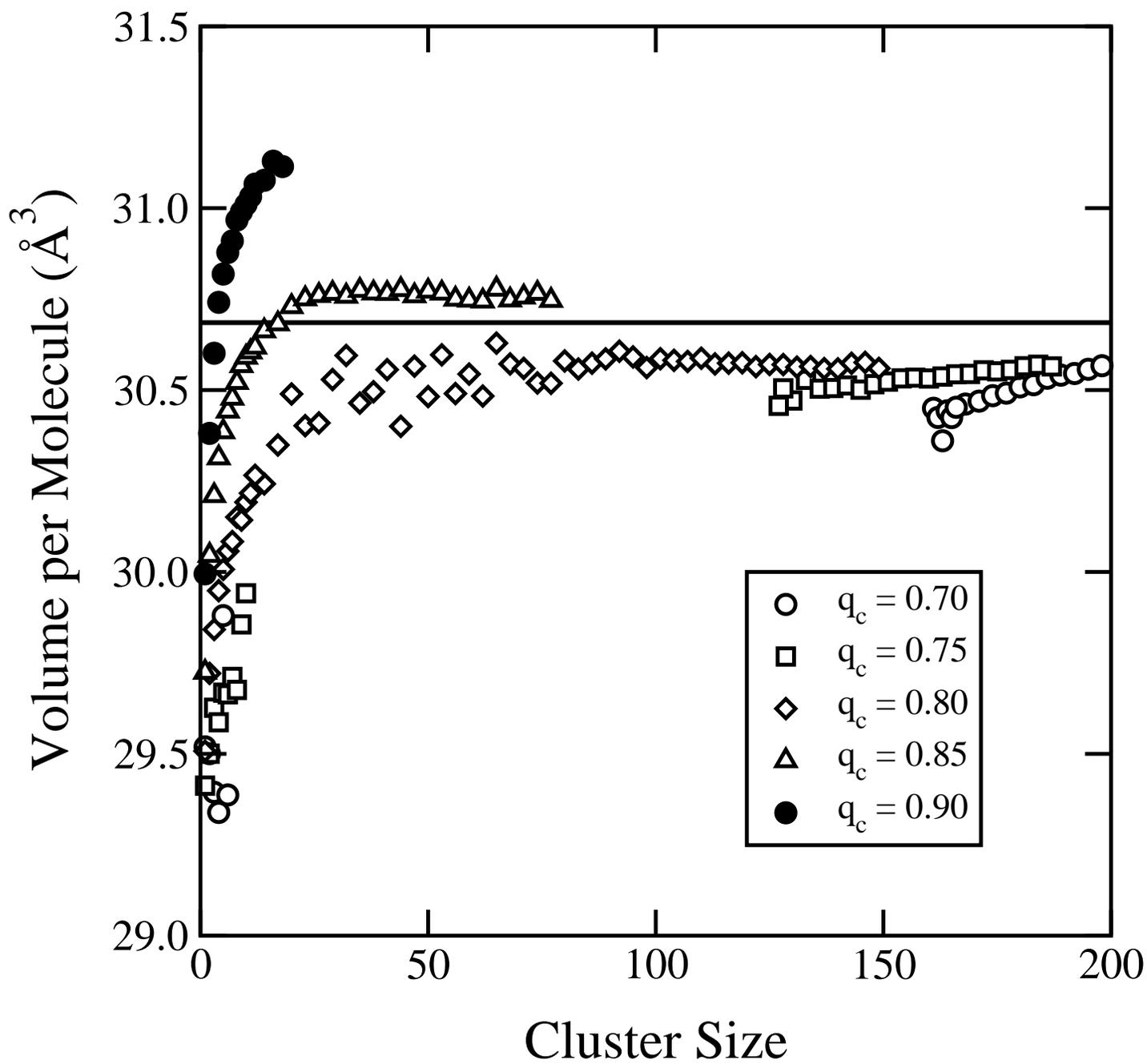

Figure 2a. Errington, Debenedetti, and Torquato

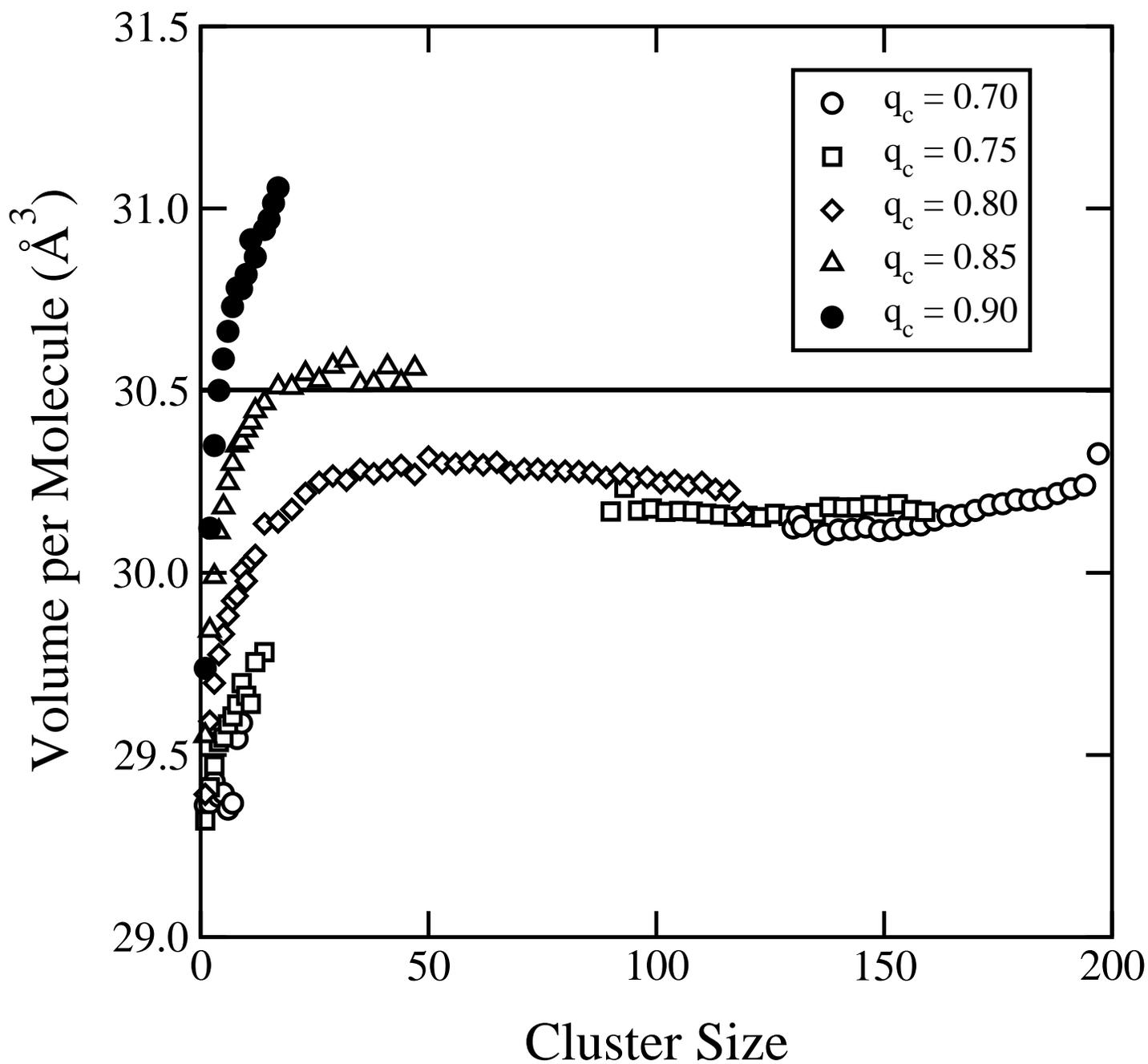

Figure 2b. Errington, Debenedetti, and Torquato

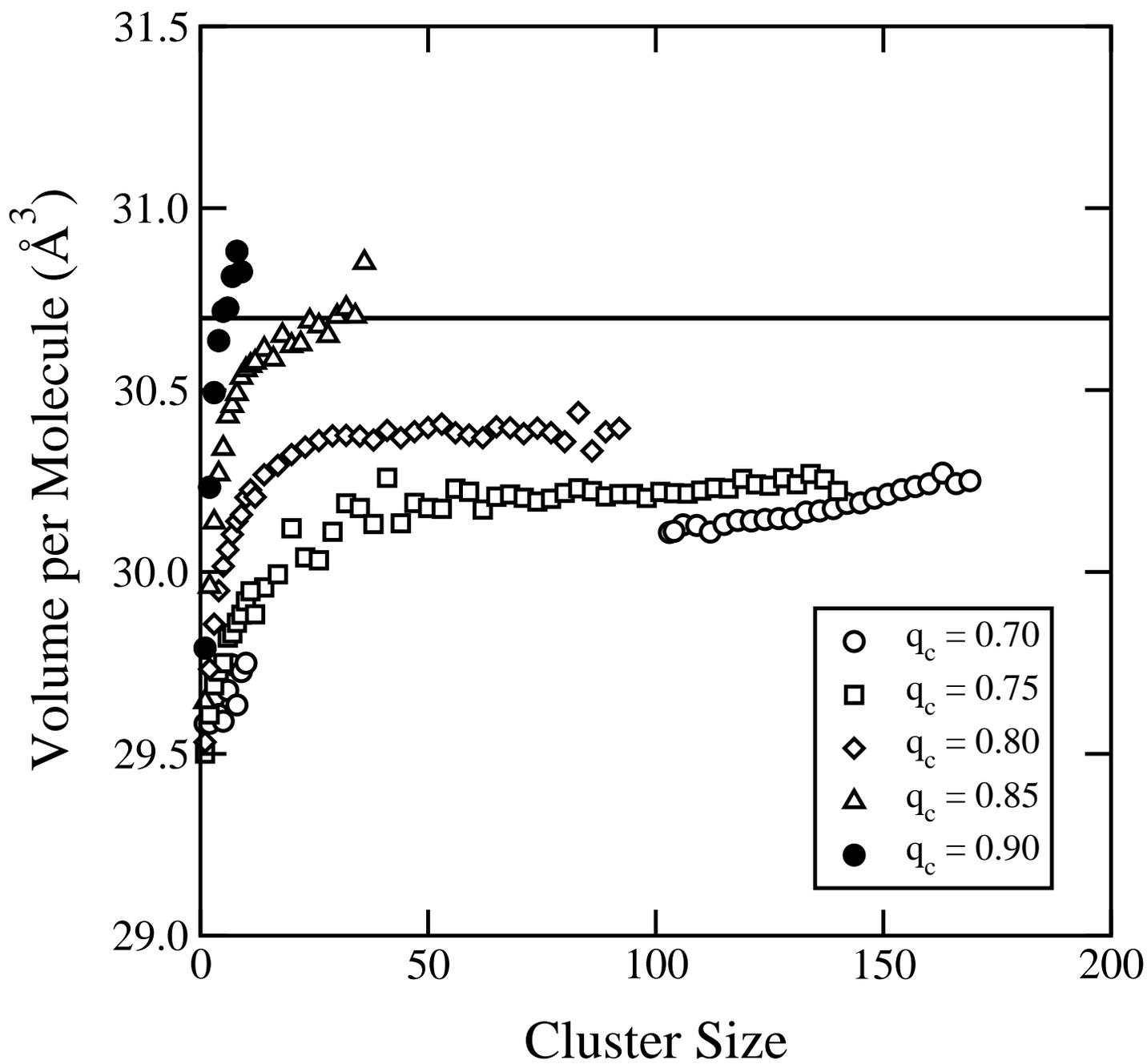

Figure 2c. Errington, Debenedetti, and Torquato

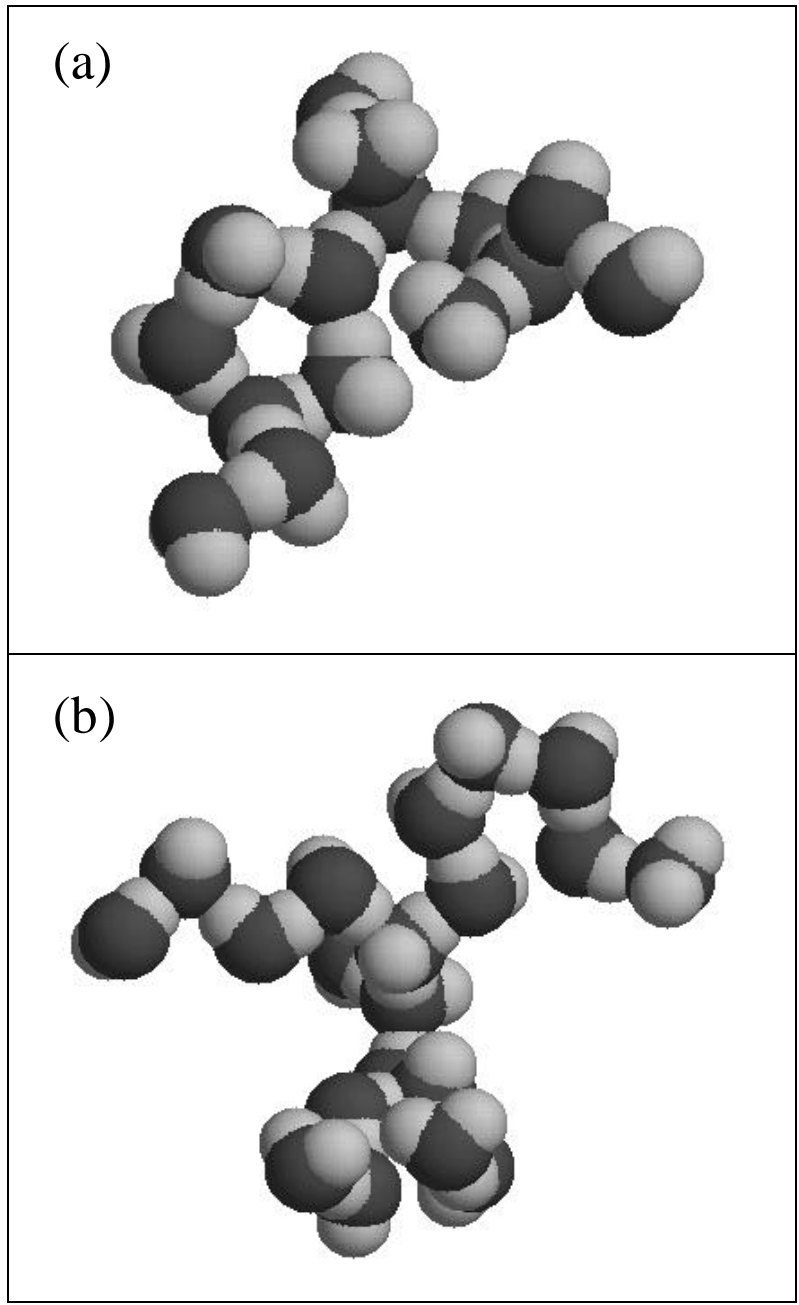

Figure 3. Errington, Debenedetti, and Torquato

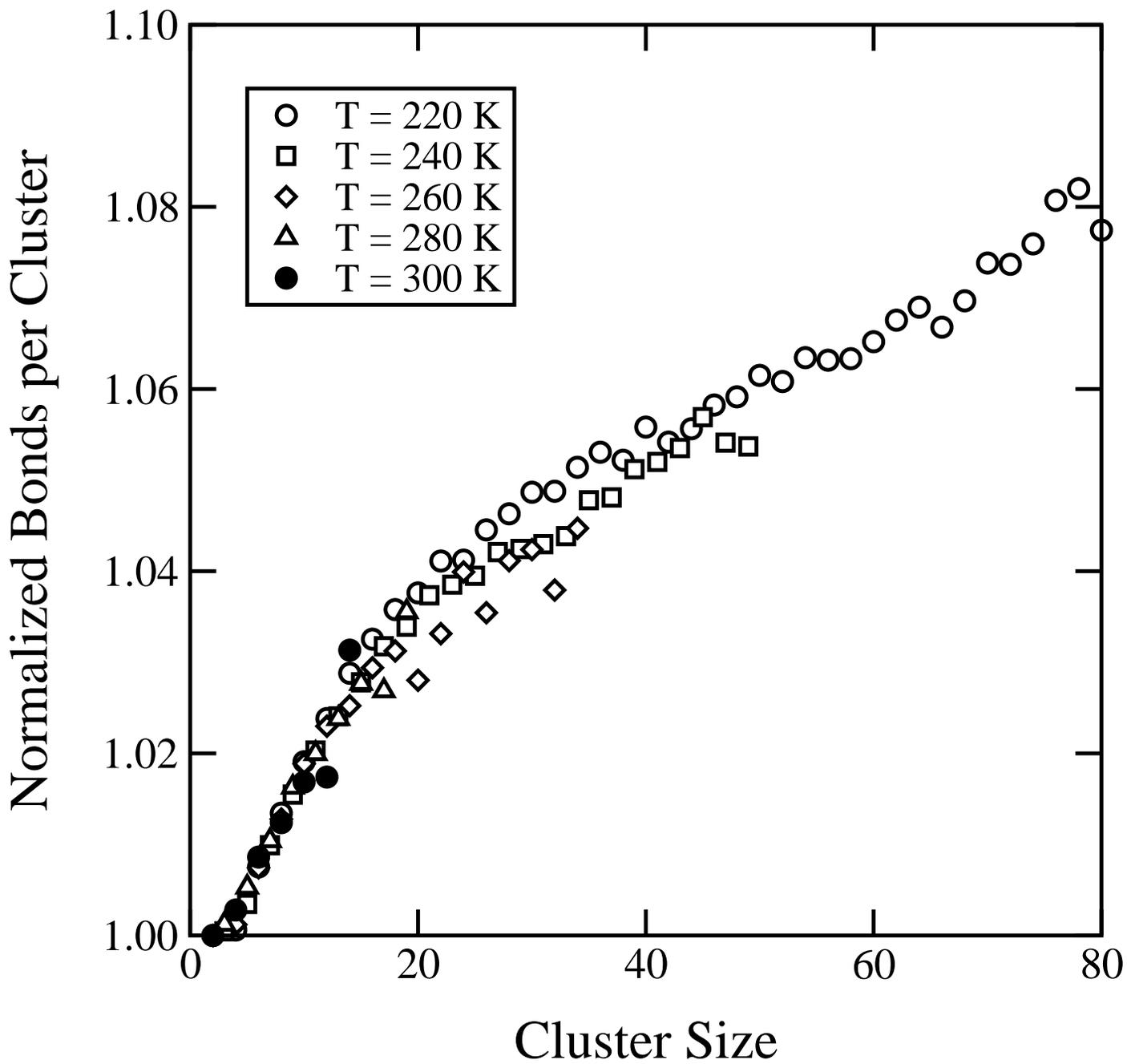

Figure 4. Errington, Debenedetti, and Torquato

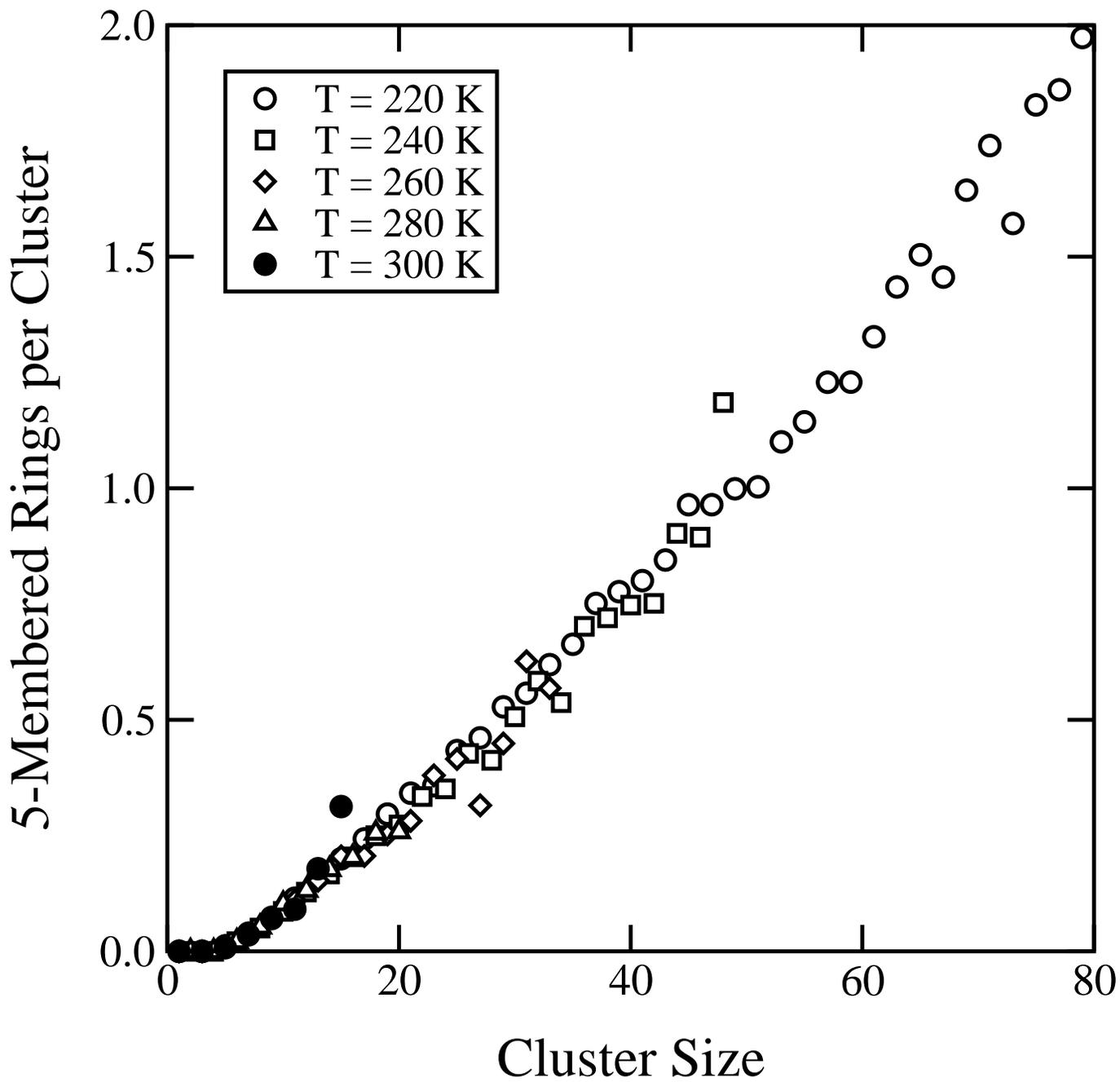

Figure 5. Errington, Debenedetti, and Torquato

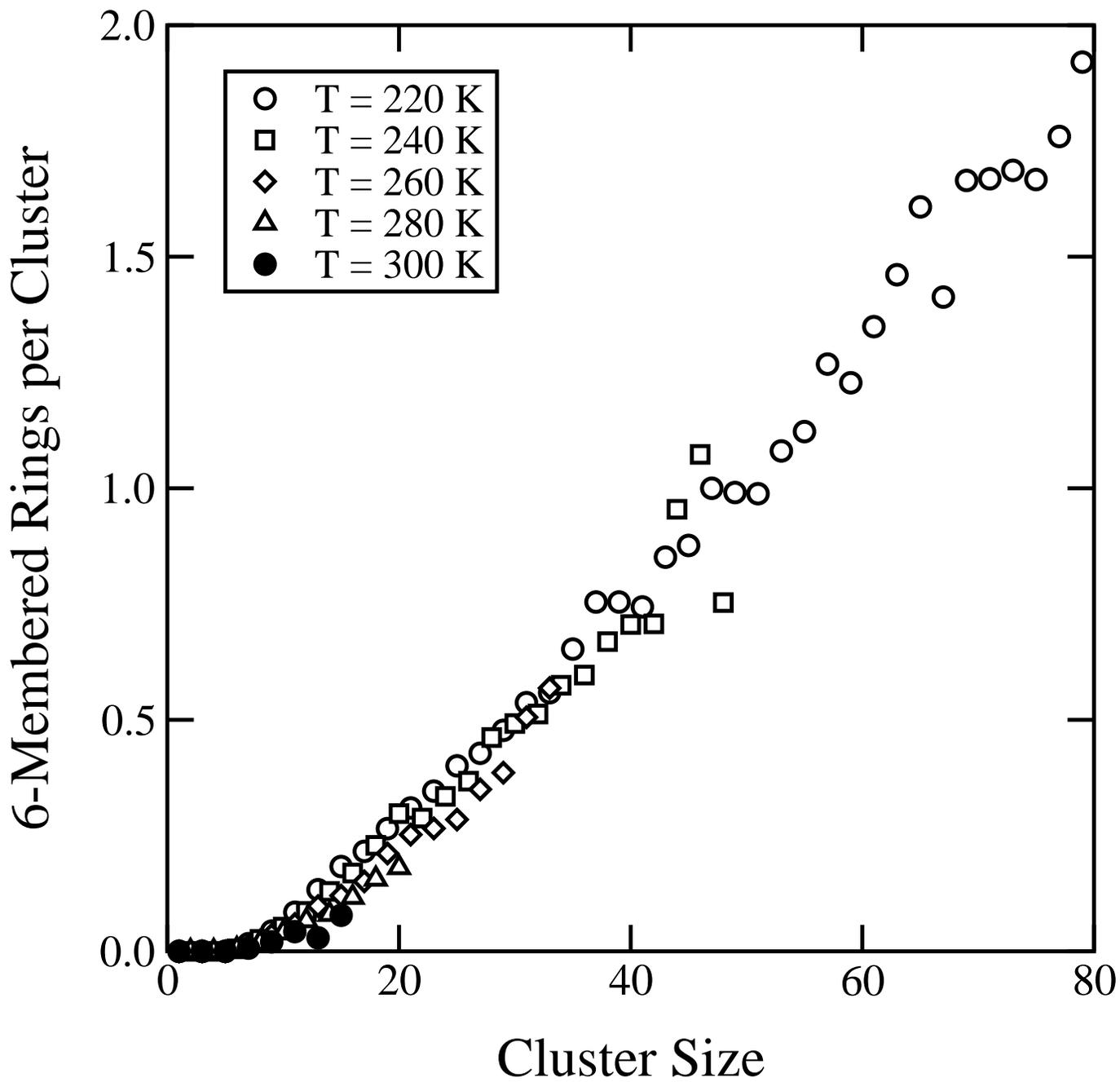

Figure 6. Errington, Debenedetti, and Torquato